\documentclass[aps,pre,showpacs,twocolumn]{revtex4}
\usepackage{bm}
\usepackage{graphicx}
\bibstyle{approve.bib}
\usepackage{amssymb}
\usepackage{epstopdf}
\usepackage{color}
\usepackage{amsmath}
\usepackage{esint}
\usepackage{gensymb}

\newcommand{\be}{\begin{equation}}
\newcommand{\ee}{\end{equation}}
\newcommand{\bea}{\begin{eqnarray}}
\newcommand{\eea}{\end{eqnarray}}

\newcommand{\br}{\mathbf{r}}

\newcommand{\tl}{\tilde{\lambda}}

\newcommand{\bk}{\mathbf{k}}

\newcommand{\e}{\varepsilon}

\newcommand{\tv}{\tilde{v}}

\newcommand{\pa}{\parallel}

\begin{document}

\title{Electrostatics of polymer translocation events in electrolyte solutions}

\author{Sahin Buyukdagli$^{1}$\footnote{email:~\texttt{Buyukdagli@fen.bilkent.edu.tr}}  and T. Ala-Nissila$^{2,3}$\footnote{email:~\texttt{Tapio.Ala-Nissila@aalto.fi}}}
\affiliation{$^{1}$Department of Physics, Bilkent University, Ankara 06800, Turkey\\
$^{2}$Department of Applied Physics and COMP Center of Excellence, Aalto University School of Science, P.O. Box 11000, FI-00076 Aalto, Espoo, Finland\\
$^{3}$Department of Physics, Brown University, Providence, Box 1843, RI 02912-1843, U.S.A.}
\date{March 18, 2016}

\begin{abstract}
We develop an analytical theory that accounts for the image and surface charge interactions between a charged dielectric membrane and a DNA molecule translocating through the membrane. Translocation events through neutral carbon-based membranes are driven by a competition between the repulsive DNA-image-charge interactions and the attractive coupling between the DNA
segments on the {\it trans} and the {\it cis} sides of the membrane. The latter effect is induced by the reduction of the coupling by the dielectric membrane. In strong salt solutions where the repulsive image-charge effects dominate the attractive {\it trans-cis} coupling, the DNA molecule encounters a translocation barrier of $\approx10$ $k_BT$. In dilute electrolytes, the {\it trans-cis} coupling takes over image-charge forces and the membrane becomes a metastable attraction point that can trap translocating polymers over long time intervals. This mechanism can be used in translocation experiments in order to control the DNA motion by tuning the salt concentration of the solution.
\end{abstract}

\date{\today}
\maketitle

\begin{figure*}
\includegraphics[width=.49\linewidth]{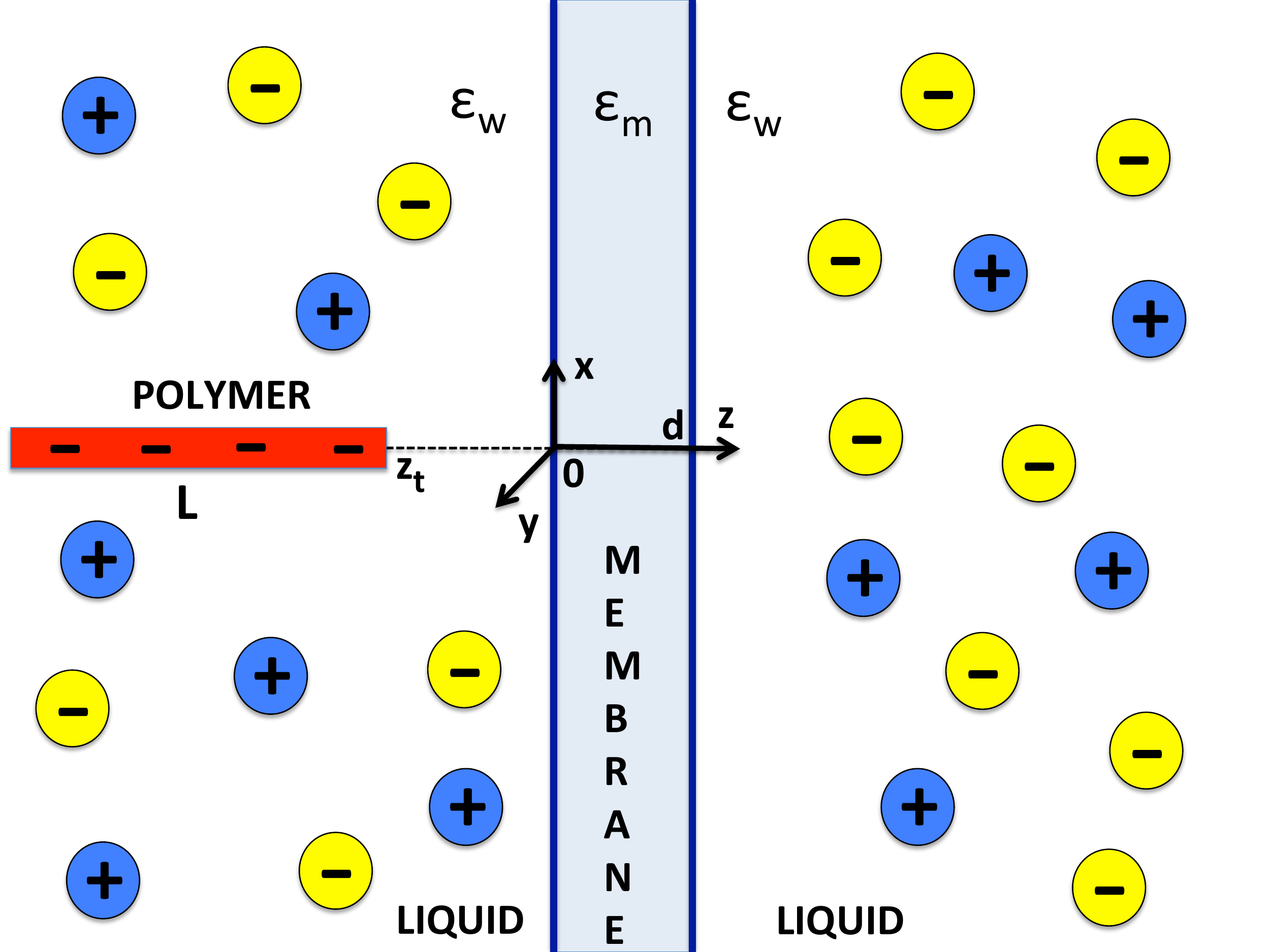}
\includegraphics[width=.49\linewidth]{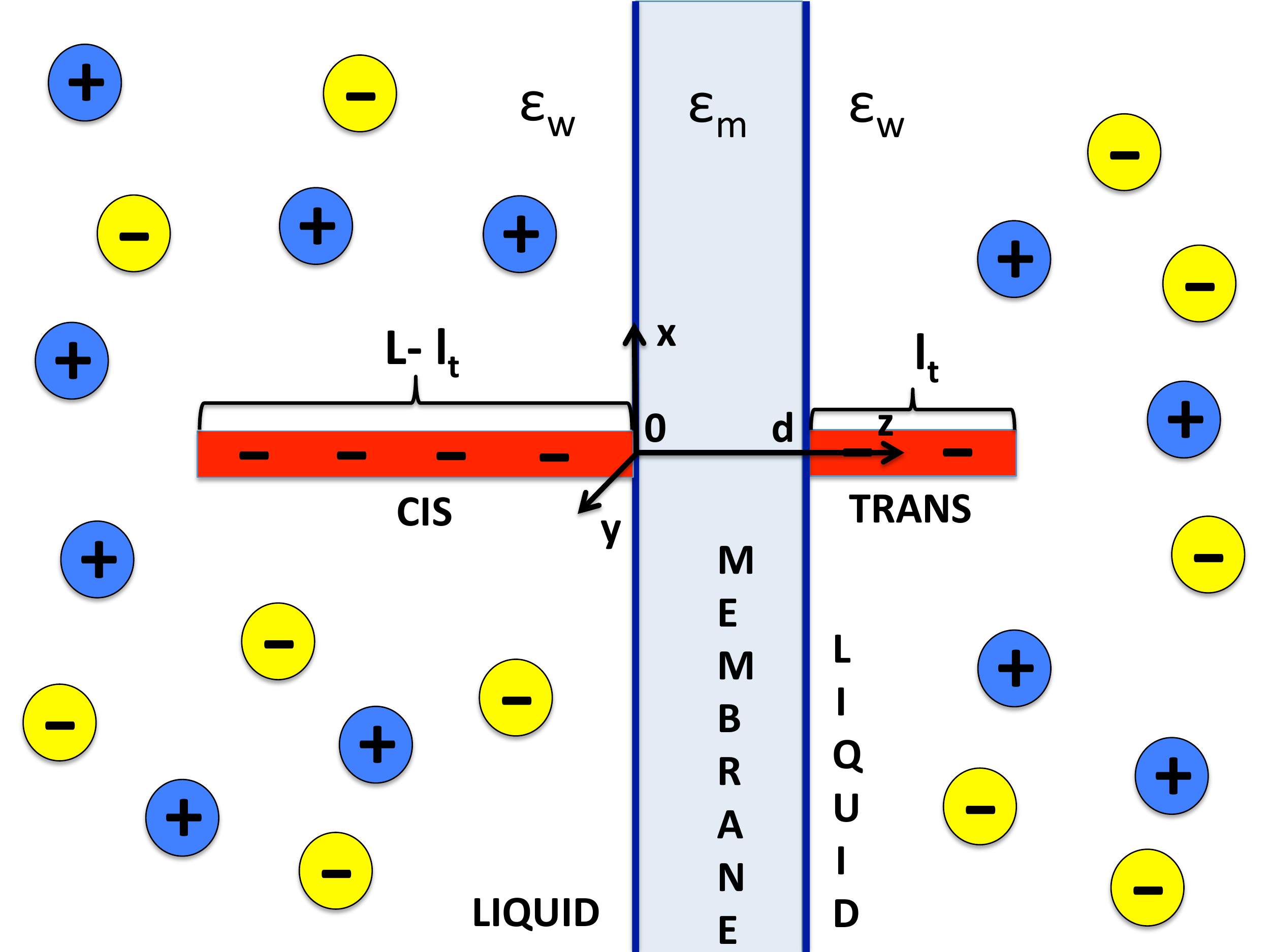}
\caption{(Color online) A schematic representation of the membrane with dielectric permittivity $\e_m$ and thickness $d$, and the ds-DNA molecule with length $L \gg d$, immersed in a monovalent electrolyte solution. Left panel: Approach of the DNA towards the membrane at distance $z_t<0$ from the membrane surface on the {\it cis} side. Right panel: Translocation of the DNA from the \textit{cis} to the \textit{trans} side quantified by the length of the translocated part of the rod $l_t$.}
\label{fig1}
\end{figure*}

\section{Introduction}

Macromolecular interactions are the driving force behind many biotechnological applications. Among them, the electrophoretic translocation of DNA molecules through membrane nanopores has recently attracted a huge amount of attention. The sequencing method proposed by Kasianowicz {\it et al.}~\cite{e1} aims at decoding the genetic content of a translocating DNA sequence via the variations of the ion flux through the nanopore. In order to facilitate the mapping between the ionic current signal and the genetic information, extensive experimental \cite{e2,e3,e4,e5,e6,e7,e8,e9,e10,e11,e12} and 
theoretical \cite{theory} efforts have been taken 
during the past three decades. Despite the progress achieved so far, there are still many outstanding 
problems in polymer translocation.

A central feature of sequence reading during translocation is the control over the DNA motion. It was recently shown that a mapping between the genetic content and the ionic current fluctuations can be established exclusively for the translocation events with the longest lifetime~\cite{e5}. Thus,  improving the accuracy of this method necessitates the reduction of the DNA translocation velocity. Achieving this goal by reducing the externally applied electric field is clearly not optimal since the precision of this approach also depends on the strength of the ionic current signal. Thus, it would be desirable to control the DNA motion 
independently of the external electric field. To this end, quantitative characterization of the interactions between the DNA and its surrounding medium becomes essential.

The complexity of the translocation problem stems from the complicated entropic, hydrodynamic, and electrostatic interactions between the polymer, the solvent molecules, the ions, and the translocated membrane. Previous theoretical and numerical investigations of translocation processes have mainly focused on entropic issues related to the flexibility of the polymer and its steric interactions with the nanopore~\cite{n1,n2,n3,n4}. Considering that the highly charged DNA molecules with line charge 
density $\lambda=2\;e/(0.34\;\mbox{nm})$ are strongly coupled to the dielectric membrane and the mobile ions in the solution, neglecting electrostatic interactions is clearly a drastic approximation. Important steps in this direction have been taken by Ghosal~\cite{Ghosal2006,Ghosal2007} and Muthukumar~\cite{muthu1,muthu2}, who coupled the Stokes-level hydrodynamics with the mean-field (MF) level electrostatics of DNA. These MF formalisms provided an elegant and analytically transparent electrohydrodynamic theory of polymer translocation. 

The MF electrostatics based on the Poisson-Boltzmann (PB) equation is known to fail in the presence of multivalent ions or any dielectric contrast in the system. The latter fact is particularly important in polymer translocation since the artificial and biological membranes used in translocation experiments are usually made of carbon-based materials with a low dielectric permittivity $\e_m\approx2$. In view of the high solvent dielectric permittivity $\e_w\approx80$, one expects strong image-charge forces acting on the mobile ions and the portions of the translocating polyelectrolyte located inside and outside the nanopore. In order to overcome this problem, in Ref.~\cite{Buyuk2014} we introduced the first correlation-corrected electrohydrodynamic theory of polymer translocation. Within this theory that includes image-charge effects and correlations at the full one-loop level, we showed that adding multivalent counterions to the solution presents itself as an efficient way to control the DNA translocation velocity. In particular, a sufficient amount of multivalent cations can neutralise or even invert the DNA charge. This effect can in turn stop the translocating DNA or reverse its direction. It should be noted that the reversal of the electrophoretic DNA mobility has also been observed in the MD simulations of Luan and Aksimentiev~\cite{Luan2009} and in translocation experiments~\cite{soft}.

The main limitation of the beyond-MF formalism of Ref.~\cite{Buyuk2014} is that the model accounts exclusively for the portion of the DNA located inside the nanopore and neglects the portions on the trans and cis sides. This approximation is valid in describing DNA translocation through thick membranes. However, it should be noted that the thickness of the graphene-based membranes used in translocation experiments can be reduced up to $d \approx 6$ {\AA}~\cite{Garaj}. Hence, at any given time during translocation
most of the polymer segments are either on the {\it cis} or the {\it trans} side, experiencing image-charge forces induced by the dielectric contrast between the solvent and the membrane. Such forces are expected to have a strong influence on the translocation process. Motivated by this fact, in the present work we develop a polymer translocation theory that accounts for the interactions between the membrane and the segments of the DNA located outside the nanopore. In Section~\ref{for}, we calculate the grand potential of the translocating DNA through a charged dielectric membrane. The theory is an extension of the model in Ref.~\cite{Buyuk2016}, where we considered the approach of a DNA molecule towards planar membranes. In Section~\ref{sec}, we scrutinize the effect of the membrane dielectric permittivity, the electrolyte density, the polymer length, and the membrane charge on the translocation process. The approximations of the model and future extensions are discussed in the Conclusions and Summary section.

\section{Electrostatic Translocation model}
\label{for}

\subsection{General formalism}

We present here an electrostatic model of a charged polymer translocating through a membrane of thickness $d$ and dielectric permittivity $\e_m$. The model is depicted in Fig.~\ref{fig1}. The left panel illustrates the approach phase of the DNA that was scrutinized in Ref.~\cite{Buyuk2016}. In the present work, we extend the model by including the most crucial phase of the DNA transport, namely the actual translocation process depicted by the right panel of Fig.~\ref{fig1}.  The membrane and the polymer are both immersed in a symmetric electrolyte with two monovalent ionic species $q=\pm1$, bulk concentration $\rho_b$, and dielectric permittivity $\e_w=80$. The electrolyte located at $z<0$ and $z>d$ is assumed to be thermalized at ambient temperature $T=300$ K.  We also note that dielectric permittivities are expressed in units of the air permittivity, i.e. we set $\e_0=1$.

The polymer of length $L$ is modeled as a rigid line charge perpendicular to the membrane at all times. 
The polymer charge density function is
\be\label{pc}
\sigma_p(\br)=-\lambda\;\delta\left(\br_\pa\right)\;g(z),
\ee
where $\lambda=2\;e/(0.34\;\mbox{nm})$ stands for the bare line charge density of the ds-DNA. Furthermore, $\br_\pa$  is the vector indicating the position of any point in the $x-y$ plane that coincides with the lateral membrane surface, and $g(z)$ is the polymer structure factor along the $z$ axis. In the most general situation, we assume that the walls of the membrane are uniformly charged, with the membrane charge density function
\be\label{mc}
\sigma_m(\br)=\sigma_m\left[\delta(z)+\delta(z-d)\right].
\ee

In Ref.~\cite{Buyuk2016} it was shown that in the case of a ds-DNA approaching a charged dielectric membrane, the electrostatic polymer grand potential associated with the presence of the membrane is composed of two contributions. These are the polymer-self energy $\Delta\Omega_{\rm pol}$  induced by polymer-image charge interactions  and the polymer-membrane charge interaction $\Omega_{\rm pm}$,
\be
\label{eq0}
\Delta\Omega_{\rm tot}=\Delta\Omega_{\rm pol}+\Omega_{\rm pm}.
\ee
We note that in Eq.~(\ref{eq0}), the additivity of these two contributions results from the Debye-Huckel (DH) level evaluation of the polymer grand potential. Moreover, since Eq.~(\ref{eq0}) is independent of the geometry, this equality is also valid for the translocation phase. In Appendix~\ref{a1}, we show that in the general case, the polymer self-energy is given by the integral 
\be
\label{eq1}
\frac{\Delta\Omega_{\rm pol}}{k_BT}=\lambda^2\int_0^\infty\frac{\mathrm{d}kk}{4\pi}\iint_{-\infty}^{+\infty}\mathrm{dz}\mathrm{dz'}g(z)\delta\tv_{\rm DH}(z,z')g(z'),
\ee
with the function $\delta\tv_{\rm DH}(z,z')$  defined by Eq.~(\ref{eqA11}) is the part of the Fourier-transformed electrostatic Green's function solely due to the presence of the dielectric membrane. Then, the second term of Eq.~(\ref{eq0}) 
which takes into account the interaction between the polymer and the membrane charges reads
\be
\label{eq1II}
\Omega_{\rm pm}=k_BT\int\mathrm{d}\br\sigma_p(\br)\psi_m(\br),
\ee
where the function $\psi_m(\br)$ is the electrostatic potential induced by the interfacial charge distribution on the pore walls. Taking into account the planar symmetry $\psi_m(\br)=\psi_m(z)$ and inserting the polymer charge distribution function of Eq. (\ref{pc})
in Eq.~(\ref{eq1II}), the polymer-membrane coupling energy takes the form
\be
\label{eq1III}
\Omega_{\rm pm}=-k_BT\lambda\int_{-\infty}^{\infty}\mathrm{d}zg(z)\psi_m(z).
\ee
In Eq.~(\ref{eq1III}), the electrostatic potential associated with the membrane charge is the solution of the linearised Poisson-Boltzmann (PB) equation
\be
\label{eq1IV}
\left[\partial_z\e(z)\partial_z-\e_w\kappa^2\theta(-z)\theta(z-d)\right]\psi_m(z)=-4\pi\ell_B\e_w\sigma_m(\br).
\ee
In Eq.~(\ref{eq1IV}), the dielectric permittivity function $\e(z)$ is given by Eq.~(\ref{eqA4}) of Appendix~\ref{a1}. We also 
introduced the Bjerrum length $\ell_B=e^2/(4\pi\e_wk_BT)\approx 7$ {\AA} and  the DH screening parameter $\kappa^2=8\pi q^2\ell_B\rho_b$. Moreover, the product of the Heaviside step functions on the r.h.s. of Eq.~(\ref{eq1IV}) takes into account the absence of charges in the membrane. The solution to Eq.~(\ref{eq1IV}) satisfying the continuity of the potential $\psi_m(z)$ and the displacement field $\e(z)\psi'_m(z)$ at $z=0$ and $z=d$ reads
\bea\label{eq1V}
\psi_m(z)&=&\frac{2}{\kappa\mu}\left[e^{\kappa z}\theta(-z)+\theta(z)\theta(d-z)\right.\nonumber\\
&&\left.\hspace{6mm}+e^{-\kappa (d-z)}\theta(z-d)\right],
\eea
with the Gouy-Chapman length $\mu=1/(2\pi\ell_B\sigma_m)$ characterizing the thickness of the counterion layer bound to the membrane wall.

We note that the polymer grand potential of Eq. (\ref{eq1}) and the polymer-membrane charge interaction of Eq. (\ref{eq1III}) exhibit a quadratic dependence on the bare DNA charge $\lambda$ and a linear dependence on the membrane charge $\sigma_m$,
respectively. This stems from the present DH approximation made for the sake of analytical simplicity.  As we will consider only weakly charged membranes, the DH approximation is legitimate in the calculation of the potential induced by the membrane charge. However, in the presence of strongly charged polyelectrolytes such as ds-DNA molecules in the solution, the DH formalism is known to overestimate the strength of electrostatic interactions. Thus, in order to overcome the DH approximation, we will opt for a variational charge renormalisation approach developed in Ref.~\cite{netzvar}. From now on, we will replace the bare polymer charge density $\lambda$ by an effective charge density
\be
\label{eq1VI}
\tl=\eta\lambda,
\ee
where $\eta$ stands for the polymer charge renormalisation factor in a bulk electrolyte. 

\begin{figure}
\includegraphics[width=1.05\linewidth]{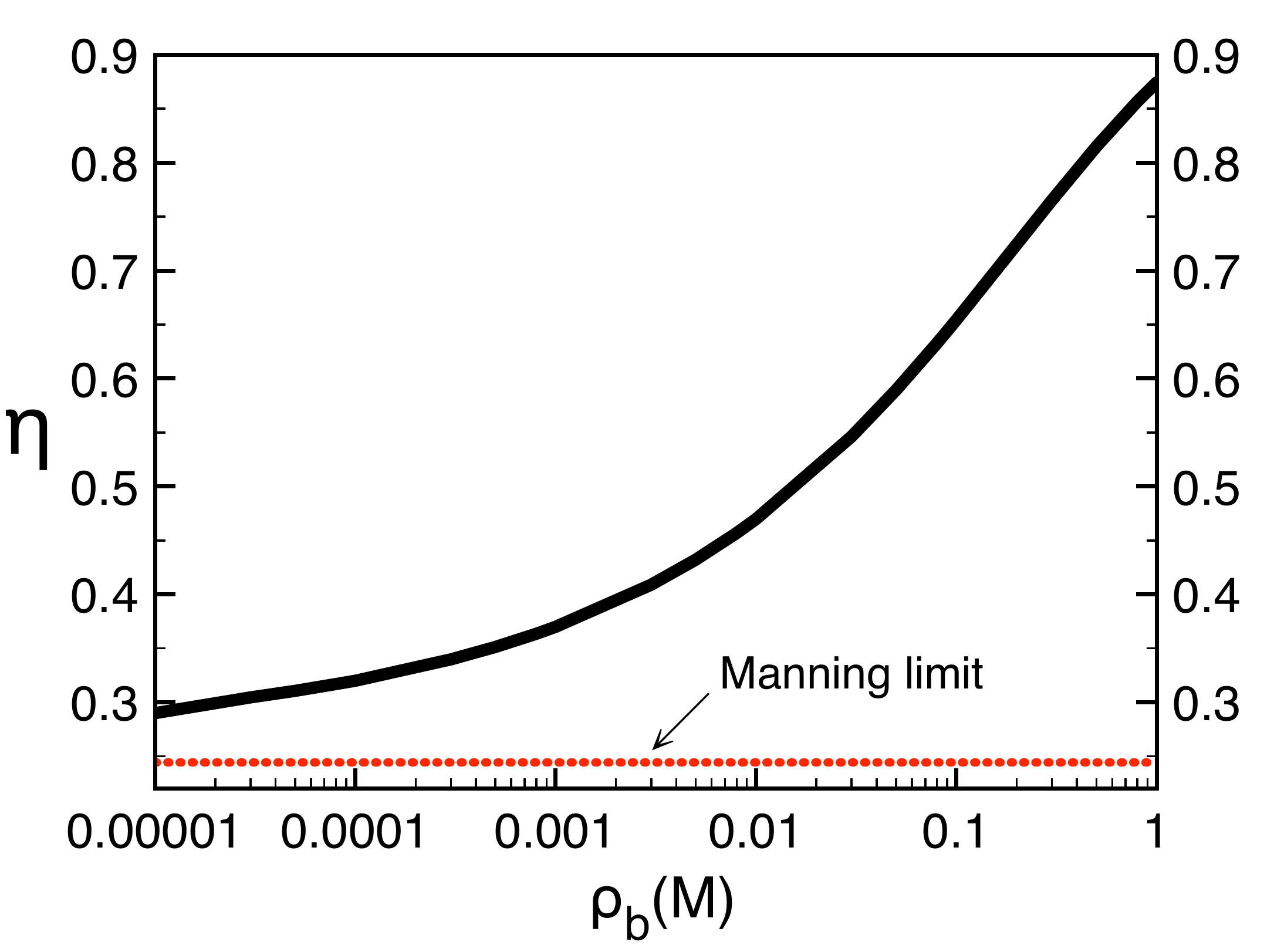}
\caption{(Color online) Charge renormalisation factor of a cylindrical ds-DNA molecule of infinite length located in a bulk electrolyte against the salt concentration. The molecule has radius $a=1$ nm and the bare line charge density is $\lambda=2\;e/(0.34\;\mbox{nm})$. The dashed red curve displays the Manning limit $\rho_b\to0$ where $\eta=1/(\ell_B\lambda)\approx0.24$.}
\label{fig1II}
\end{figure}

We will briefly describe the application of the renormalisation procedure of Ref.~\cite{netzvar} to our system. The approach consists in inserting the rescaled electrostatic potential $\eta\psi_p(r)$ into the MF-level electrostatic grand potential, with the bare potential $\psi_p(r)$ which is the solution of the linear PB equation for a charged cylinder immersed in a bulk electrolyte~\cite{netzvar},
\be\label{eq1VII}
\psi_p(r)=\frac{2\ell_B\lambda}{\kappa a}\frac{\mathrm{K}_0(\kappa r)}{\mathrm{K}_1(\kappa r)}.
\ee
In Eq.~(\ref{eq1VII}), $\mathrm{K}_0(x)$ and $\mathrm{K}_1(x)$ are the modified Bessel functions~\cite{math}, $a=1$ nm stands for the ds-DNA radius, and $r$ is the distance from the axis of symmetry of the molecule. Optimizing the resulting variational grand potential with respect to the variational charge renormalisation factor $\eta$, we obtain the integral equation
\bea
\label{eq1VIII}
&&2(1-\eta)\ell_B\lambda \psi_p(\kappa a)\\
&&+\kappa^2\int_{a}^{\infty}\mathrm{d}rr\left\{\eta\psi^2_p(r)-\psi_p(r)\sinh\left[\eta\psi_p(r)\right]\right\}=0.\nonumber
\eea
In Fig.~\ref{fig1II}, we illustrate the numerical solution of Eq.~(\ref{eq1VIII}) versus the bulk salt density. Decreasing the ion density from $\rho_b=1.0$ M to $10^{-5}$ M, the renormalisation factor drops from $\eta\approx0.9$ to $\eta\approx0.3$. This behaviour reflects the reduction of the bare DNA charge by the cations bound to the polymer. Moreover, as shown in Ref.~\cite{netzvar}, approaching the pure solvent limit  $\rho_b\to0$ this curve converges logarithmically slowly to the \textit{Manning limit}
\be\label{man}
\tl=\frac{1}{\ell_B},
\ee
or $\eta=1/(\ell_B\lambda)\approx0.24$. The plot also indicates that in the regime $\rho_b\lesssim0.1$ M, where the factor $\eta$ strongly deviates from unity, the DH approximation that assumes $\eta=1$ significantly overestimates the net DNA charge density. In terms of the renormalised charge from Eq. (\ref{eq1VI}), obtained from the solution of Eq.~(\ref{eq1VIII}), we next calculate the explicit form of the polymer grand potential in the approach and translocation phases.

\subsection{Polymer grand potential in the approach and translocation phases}

In the case of a polymer of length $L$ approaching the membrane from left, with the right end located at the distance $z_t\leq0$ from the membrane surface (see the left panel of Fig.~\ref{fig1}), the structure factor reads
\be\label{eq2}
g(z)=\theta(-z)\theta(z_t-z)\theta(z-z_t+L).
\ee
Inserting this into Eq.~(\ref{eq1}) with the Green's function~(\ref{eqA6}) and carrying out the spatial integrals, the self-energy of the approaching polymer follows as
\bea
\label{eq3}
\frac{\Delta\Omega_{\rm pol}(z_t)}{k_BT}&=&\frac{\ell_B\tl^2}{2}\int_0^\infty\frac{\mathrm{d}kk}{p^3}\frac{\Delta\left(1-e^{-2kd}\right)}{1-\Delta^2e^{-2kd}}\\
&&\hspace{1.6cm}\times\left(1-e^{-pL}\right)^2e^{2pz_t}.\nonumber
\eea
In Eq.~(\ref{eq3}), we defined the screening function $p=\sqrt{k^2+\kappa^2}$ and the dielectric discontinuity function $\Delta=(\e_wp-\e_mk)/(\e_wp+\e_mk)$. Substituting now the membrane potential of Eq. (\ref{eq1V}) into Eq.~(\ref{eq1III}) together with the structure factor of Eq. (\ref{eq2}), the polymer-membrane interaction potential associated with the approach phase takes the form
\be\label{eq3II}
\frac{\Omega_{\rm pm}(z_t)}{k_BT}=-\frac{2Q_{\rm eff}(L)}{\mu\kappa}e^{\kappa z_t},
\ee
where we introduced the effective charge of a polymer of length $L$
\be\label{eq3III}
Q_{\rm eff}(L)=\tl L\frac{1-e^{-\kappa L}}{\kappa L}.
\ee

Equations~(\ref{eq2})-(\ref{eq3III}) characterizing the approach phase have been derived in Ref.~\cite{Buyuk2016} within the pure DH limit $\eta=1$ (i.e. $\tl=\lambda$). The equations that will be introduced in the rest of the manuscript are the original results. Next, we calculate the electrostatic grand potential of the polyelectrolyte translocating through the membrane. This configuration is depicted in the right panel of Fig.~\ref{fig1}.  Since the presence of a finite length pore breaks the planar geometry of the system and complicates the theory, we simplify the model by neglecting the part of the polymer located inside the pore~\cite{lev} ($L, l_t \gg d$). Within this simplified model, the polymer of total length $L$ is composed of a section of length $l_t$ on the {\it trans} side and the other section with length $L-l_t$ on the {\it cis} side. For this configuration, the charge structure factor is given by
\be\label{eq8}
g(z)=\theta(-z)\theta(z-L+l_t)+\theta(z-d)\theta(d+l_t-z).
\ee
Inserting the function~(\ref{eq8}) into Eq.~(\ref{eq1}), the polymer self-energy splits into two parts,
\be\label{eq9}
\Delta\Omega_{\rm pol}(l_t)=\Delta\Omega_{\rm intra}(l_t)+\Delta\Omega_{\rm inter}(l_t),
\ee
where the first contribution resulting from the self-interaction between the parts of the polymer on the {\it cis}
and the {\it trans} sides is given by
\bea\label{eq10}
\frac{\Delta\Omega_{\rm intra}(l_t)}{k_BT}&=&\tl^2\int_0^\infty\frac{\mathrm{d}kk}{4\pi}\left\{\int_{-L+l_t}^{0}\mathrm{dz}\int_{-L+l_t}^{0}\mathrm{dz'}\right.\\
&&\hspace{.7cm}\left.+\int_{d}^{d+l_t}\mathrm{dz}\int_{d}^{d+l_t}\mathrm{dz'}\right\}\delta\tv_{\rm DH}(z,z'),\nonumber
\eea
and the interaction between the separate {\it cis} and {\it trans} portions reads
\bea\label{eq11}
\frac{\Delta\Omega_{\rm inter}(l_t)}{k_BT}&=&\tl^2\int_0^\infty\frac{\mathrm{d}kk}{2\pi}\int_{-L+l_t}^{0}\mathrm{dz}\int_{d}^{d+l_t}\mathrm{dz'}\delta\tv_{\rm DH}(z,z').\nonumber\\
\eea
By substituting the Green's functions~(\ref{eqA6})-(\ref{eqA8}) into Eqs.~(\ref{eq10})-(\ref{eq11}), we find 
that the polymer self-energy $\Delta\Omega_{\rm intra}(l_t)$ and the {\it trans-cis} coupling energy $\Delta\Omega_{\rm inter}(l_t)$ mediated exclusively by the membrane read
\bea
\label{eq13}
\frac{\Delta\Omega_{\rm intra}(l_t)}{k_BT}&=&\frac{\ell_B\tl^2}{2}\int_0^\infty\frac{\mathrm{d}kk}{p^3}\frac{\Delta\left(1-e^{-2kd}\right)}{1-\Delta^2e^{-2kd}}\\
&&\hspace{.5cm}\times\left\{\left[1-e^{-pl_t}\right]^2+\left[1-e^{-p(L-l_t)}\right]^2\right\};\nonumber\\
\label{eq14}
\frac{\Delta\Omega_{\rm inter}(l_t)}{k_BT}&=&\ell_B\tl^2\int_0^\infty\frac{\mathrm{d}kk}{p^3}\left\{\frac{\left(1-\Delta^2\right)e^{(p-k)d}}{1-\Delta^2e^{-2kd}}-1\right\}\nonumber\\
&&\hspace{.5cm}\times e^{-pd}\left[1-e^{-pl_t}\right]\left[1-e^{-p(L-l_t)}\right].\nonumber\\
\eea
Finally, substituting the electrostatic potential of Eq. (\ref{eq1V}) into Eq.~(\ref{eq1III}) together with the structure factor in Eq. (\ref{eq8}), the interaction energy of the translocating polymer with the membrane charge takes the form
\be
\label{eq14II}
\frac{\Omega_{\rm pm}(l_t)}{k_BT}=-\frac{2}{\mu\kappa}\left[Q_{\rm eff}(l_t)+Q_{\rm eff}(L-l_t)\right].
\ee
In Section~\ref{sec}, we characterize the electrostatics of approaching and translocating polymers in terms of the grand potentials in Equations (\ref{eq3})-(\ref{eq3II}) and~(\ref{eq13})-(\ref{eq14II}). 

\section{Results}
\label{sec}

We investigate next the electrostatic cost for the approach and the translocation of a polymer through dielectric membranes. In Sections~\ref{memper}-\ref{lh}  where we scrutinize the effect of the membrane permittivity, the salt density, and the polymer length on the translocation energetics, we consider neutral membranes (i.e. $\sigma_m=0$). Then, in Section~\ref{char}, we focus on the effect of the membrane charge on the translocation energy of ds-DNA molecules.

\subsection{Membrane dielectric permittitivity}
\label{memper}

\begin{figure}
\includegraphics[width=1.05\linewidth]{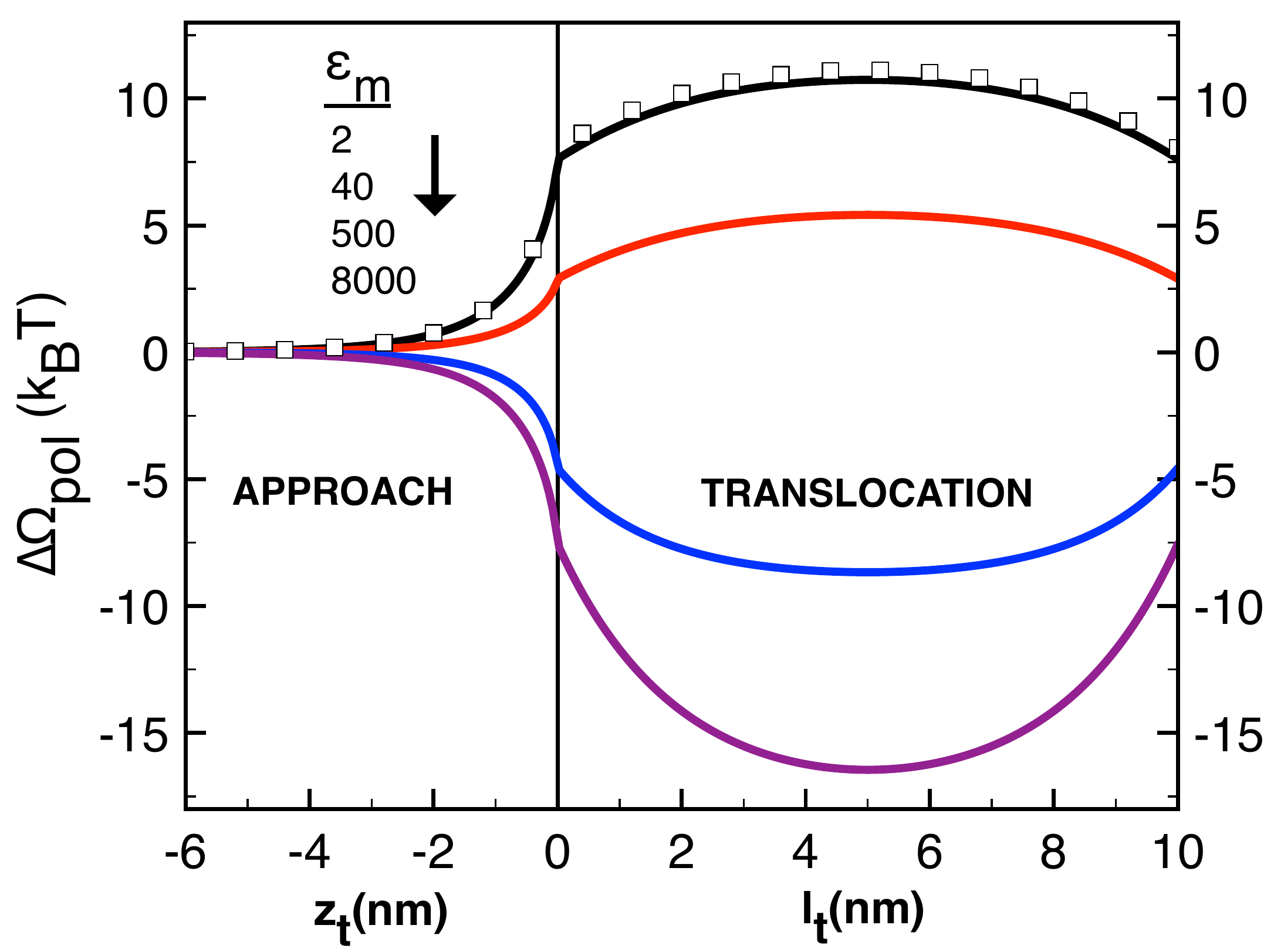}
\caption{(Color online) Grand potential of a polymer translocating a neutral membrane ($\sigma_m=0$) from Eqs.~(\ref{eq3}) and (\ref{eq13})-(\ref{eq14}) at various membrane permittivities. Salt density is $\rho_{b}=0.01$ M, membrane thickness $d=2$ nm, and polymer length $L=10$ nm. The square symbols display the low permittivity limit $\e_m\to0$ of Eqs.~(\ref{eq4}) and~(\ref{eq15})-(\ref{eq16}). }
\label{fig2}
\end{figure}

First, we consider the role played by the membrane dielectric permittivity $\e_m$ in polymer translocation through neutral membranes ($\sigma_m=0$). We plot in Fig.~\ref{fig2} the electrostatic grand potential of Eqs.~(\ref{eq3}) and (\ref{eq13})-(\ref{eq14}) for a polymer of length $L=10$ nm, a the membrane thickness of $d=2$ nm, and salt density $\rho_b=0.01$ M. The approach phase is depicted in terms of the polymer position $z_t<0$ with the (infinitesimally thin) membrane surface located at $z_t=0$. The translocation phase is in turn described in terms of the translocated length $l_t$ with $0\leq l_t\leq L$.  Although the most frequent carbon-based membranes are of low permittivity $\e_m \approx 2$, membrane engineering methods based on the inclusion of carbon structures or graphene nanoribbons (GNRs) into host matrices allow to increase the membrane permittivity up to $8000$~\cite{Dimiev,Dang}. In order to cover this permittivity range, we consider extended permittivity values. 

Figure \ref{fig2} shows that approaching the membrane of low permittivity $\e_m=2$ from the bulk region, the polymer experiences a repulsive energy that rises monotonically and reaches the value $\Delta\Omega_{\rm pol}(0)\approx8$ $k_BT$ at the membrane surface. During the translocation phase, the grand potential continues to rise and reaches its maximum value  $\Delta\Omega_{\rm pol}(l_t=L/2)\approx12$ $k_BT$ as the half of the polymer translocates. In the subsequent motion of the DNA molecule, the grand potential drops and converges to the contact value $\Delta\Omega_{\rm pol}(l_t=L)=\Delta\Omega_{\rm pol}(z_t=0)$ as 
translocation is completed. Moreover, for a lower membrane permittivity of 
$\e_m=40$ where the dielectric discontinuity weakens, the electrostatic energy barrier is lowered by a factor two. This shows that the barrier results mainly from the interaction of the DNA charges with their electrostatic images. This corresponds to the self-energy term of Eq. (\ref{eq13}) of the grand potential. The contribution of the interaction potential from Eq. (\ref{eq14}) will be investigated below. In the case of engineered membranes whose dielectric permittivity is larger than that of water (e.g. the curve for $\e_m=500$), the electrostatic grand potential of the DNA becomes negative and reaches its minimum in the half-translocated state. Thus, with the membrane permittivity exceeding the water permittivity, the membrane becomes an attraction point. In particular, at the highest dielectric permittivity value $\e_m=8000$ measured for membranes including GNRs~\cite{Dang}, the potential well reaches a significantly low value of $\Delta\Omega_{\rm pol}(l_t=L/2)\approx-17$ $k_BT$. Hence, high permittivity membranes are expected to efficiently trap translocating DNA molecules. 

At this point we should note that the charge renormalisation process introduced here allows an important correction as it lowers the approach energies evaluated in Ref.~\cite{Buyuk2016} in the DH approximation by an order of magnitude. This can be seen by comparing Fig.~\ref{fig2} of the present manuscript with the inset of Fig. 3 in Ref.~\cite{Buyuk2016}. Second, we emphasize that previous models that aimed at evaluating the electrostatic cost of DNA translocation events have focused exclusively on the energy of the translocating polymer inside the nanopore~\cite{e3,Buyuk2014}. The high values of the grand potential curves in Fig.~\ref{fig2} indicate that the contribution from the DNA segments located outside the membrane is indeed non-negligible and should play a determinant role in translocation events. This is the first important conclusion of our work. We consider next the alteration of the polymer grand potential landscapes by tuning the ion density of the liquid.

\subsection{Salt concentration} 
\label{salt}

Since salt concentration is an easily tunable parameter, we consider now the effect of salt on the electrostatic grand potential of the translocating DNA. We will focus on the most relevant case of C-based low permittivity membranes. In order to simplify the analysis, we will take the limit $\e_m=0$ where the polymer grand potential allows an analytical form. In this limit, the approach energy~(\ref{eq3}) becomes
\be\label{eq4}
\frac{\Delta\Omega_{\rm pol}(z_t)}{k_BT}=\frac{\ell_B\tl^2}{2\kappa}G(z_t),
\ee
where we introduced the adimensional auxiliary function
\bea\label{eq5}
G(z_t)&=&e^{2\kappa z_t}+e^{-2\kappa(L-z_t)}-2e^{-\kappa(L-2z_t)}\nonumber\\
&&-2\kappa z_t\;\mathrm{Ei}[2\kappa z_t]+2\kappa(L-z_t)\mathrm{Ei\left[-2\kappa(L-z_t)\right]}\nonumber\\
&&-2\kappa(L-2z_t)\mathrm{Ei\left[-\kappa(L-2z_t)\right]},
\eea
and $\mathrm{Ei}(x)$ is the exponential integral function~\cite{math}
\be
\mathrm{Ei}(x)=\int_x^\infty\mathrm{d}t\frac{e^{-t}}{t}.
\ee
Moreover, the translocating polymer free energies~(\ref{eq13})-(\ref{eq14}) take the form
\bea\label{eq15}
\frac{\Delta\Omega_{\rm intra}(l_t)}{k_BT}&=&\frac{\ell_B\tl^2}{2\kappa}H(l_t)\\
\label{eq16}
\frac{\Delta\Omega_{\rm inter}(l_t)}{k_BT}&=&-\frac{\ell_B\tl^2}{\kappa}F(l_t),
\eea
with the auxiliary functions
\bea
\label{eq17}
H(l_t)&=&\left[1-e^{-\kappa l_t}\right]^2+\left[1-e^{-\kappa(L-l_t)}\right]^2\\
&&+2\kappa l_t\left[\mathrm{Ei}(-2\kappa l_t)-\mathrm{Ei}(-\kappa l_t)\right]\nonumber\\
&&+2\kappa(L-l_t)\left\{\mathrm{Ei}\left[-2\kappa(L-l_t)\right]-\mathrm{Ei}\left[-\kappa(L-l_t)\right]\right\}\nonumber
\eea
and
\bea
\label{eq18}
F(l_t)&=&e^{-\kappa d}\left[1-e^{-\kappa l_t}\right]\left[1-e^{-\kappa(L-l_t)}\right]+\kappa d\;\mathrm{Ei}(-\kappa d)\nonumber\\
&&+\kappa(d+L)\;\mathrm{Ei}\left[-\kappa(d+L)\right]\nonumber\\
&&-\kappa(d+l_t)\;\mathrm{Ei}\left[-\kappa(d+l_t)\right]\nonumber\\
&&-\kappa(d+L-l_t)\;\mathrm{Ei}\left[-\kappa(d+L-l_t)\right].
\eea
The total polymer grand potential is obtained from Eqs.~(\ref{eq15}) and~(\ref{eq16}) via Eq.~(\ref{eq9}).
\begin{figure}
\includegraphics[width=1.05\linewidth]{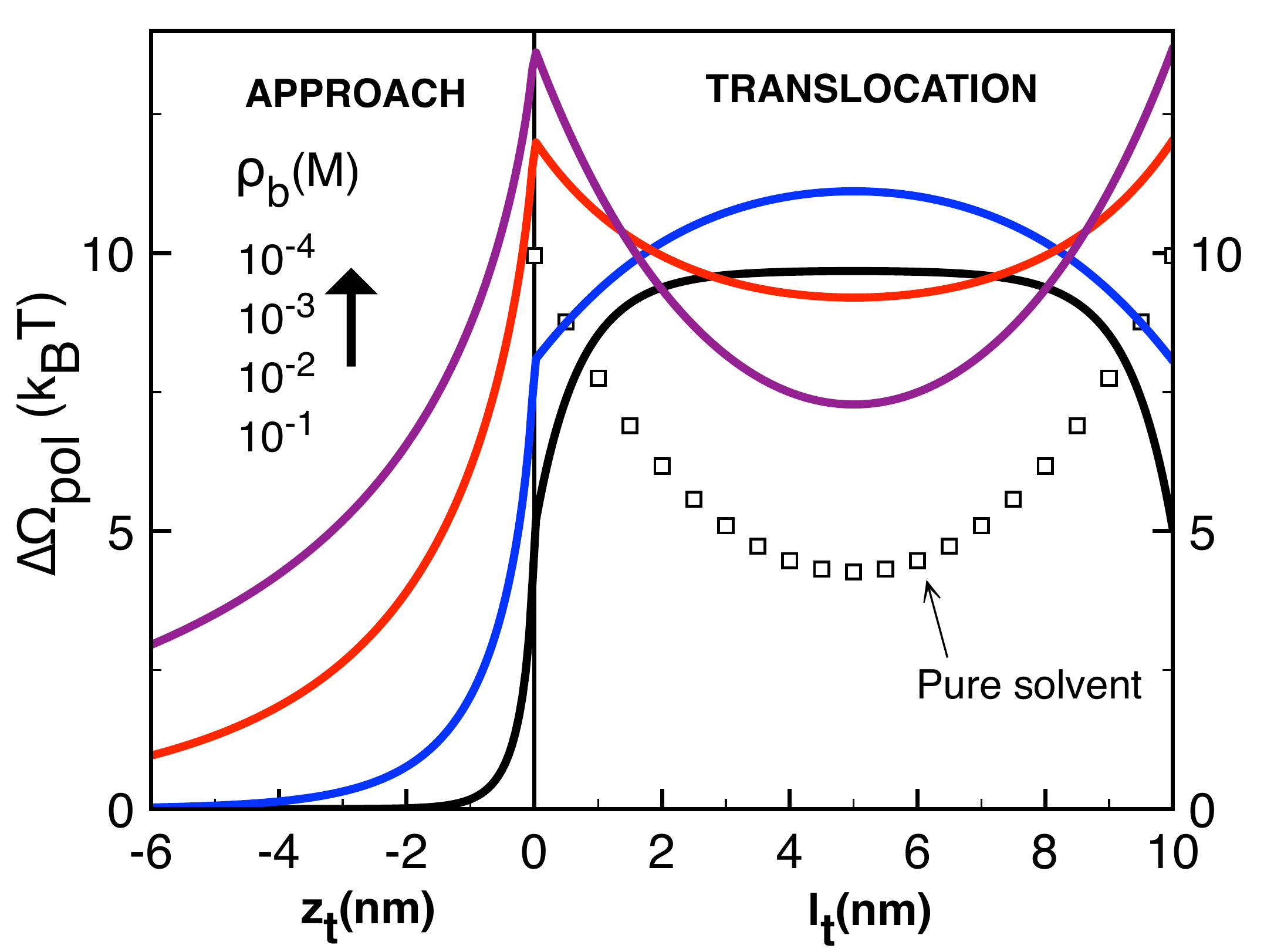}
\caption{(Color online) Polymer grand potential for $\e_m=0.0$ from Eqs.~(\ref{eq4}) and~(\ref{eq15})-(\ref{eq16}) at various salt concentrations. The square symbols show the pure solvent limit $\rho_b\to0$ of Eqs.~(\ref{eq19})-(\ref{eq20}). The remaining parameters are the same as in Fig.~\ref{fig2}.}
\label{fig3}
\end{figure}

In Fig.~\ref{fig2}, we show that the closed-form expressions~(\ref{eq4}) and~(\ref{eq15})-(\ref{eq16}) for $\e_m=0$ (open squares) accurately approximate the polymer grand potential profile at the characteristic value $\e_m=2$. Figure \ref{fig3} displays the salt dependence of the polymer grand potential. As the salt density is reduced from $\rho_b=0.1$ M to $0.01$ M, the weakened charge screening amplifies the electrostatic grand potential of the DNA during  its approach ($z_t\leq0$ ) and its translocation through the membrane ($0<l_t<L$). However, at lower ion densities, the surface potential barrier and the translocation potential exhibit opposite behaviour. Namely, with further reduction of the salt concentration, the grand potential of the approaching polymer rises monotonically for $z_t\leq0$. During the subsequent translocation phase, between $\rho_b=0.01$ M and $0.001$ M, the translocation grand potential changes its slope for $0<l_t<L$ and the barrier becomes a metastable well. With decreasing salt, the metastable minimum becomes deeper until one reaches the pure solvent limit $\rho_b\to0$ (or $\kappa\to0$) where the free energies~(\ref{eq15})-(\ref{eq16}) take with Eq.~(\ref{man}) a simple form
\bea\label{eq19}
&&\frac{\Delta\Omega_{\rm intra}(l_t)}{k_BT}=\frac{L}{\ell_B}\ln(2);\\
\label{eq20}
&&\frac{\Delta\Omega_{\rm inter}(l_t)}{k_BT}=-\frac{L}{\ell_B}\left\{\frac{d}{L}\ln\left[\frac{d(L+d)}{(d+l_t)(L+d-l_t)}\right]\right.\nonumber\\
&&\hspace{2cm}\left.+\ln\left[\frac{L+d}{L+d-l_t}\right]+\frac{l_t}{L}\ln\left[\frac{L+d-l_t}{d+l_t}\right]\right\}.\nonumber\\
\eea
The prediction of Eqs.~(\ref{eq19})-(\ref{eq20}) is illustrated in Fig.~\ref{fig3} by square symbols. The depth of the grand potential well indicates that in translocation experiments with weak electrolytes, the DNA molecule is expected to be trapped by the membrane over long time periods. This is the key result of our work. As discussed in the Introduction, in translocation experiments accurate DNA sequencing necessitates the reduction of the translocation velocity of DNA~\cite{e5}. Thus, the observed effect can be efficiently used to control the DNA velocity via alteration of the salt density in the solution.  

\begin{figure}
\includegraphics[width=1.25\linewidth]{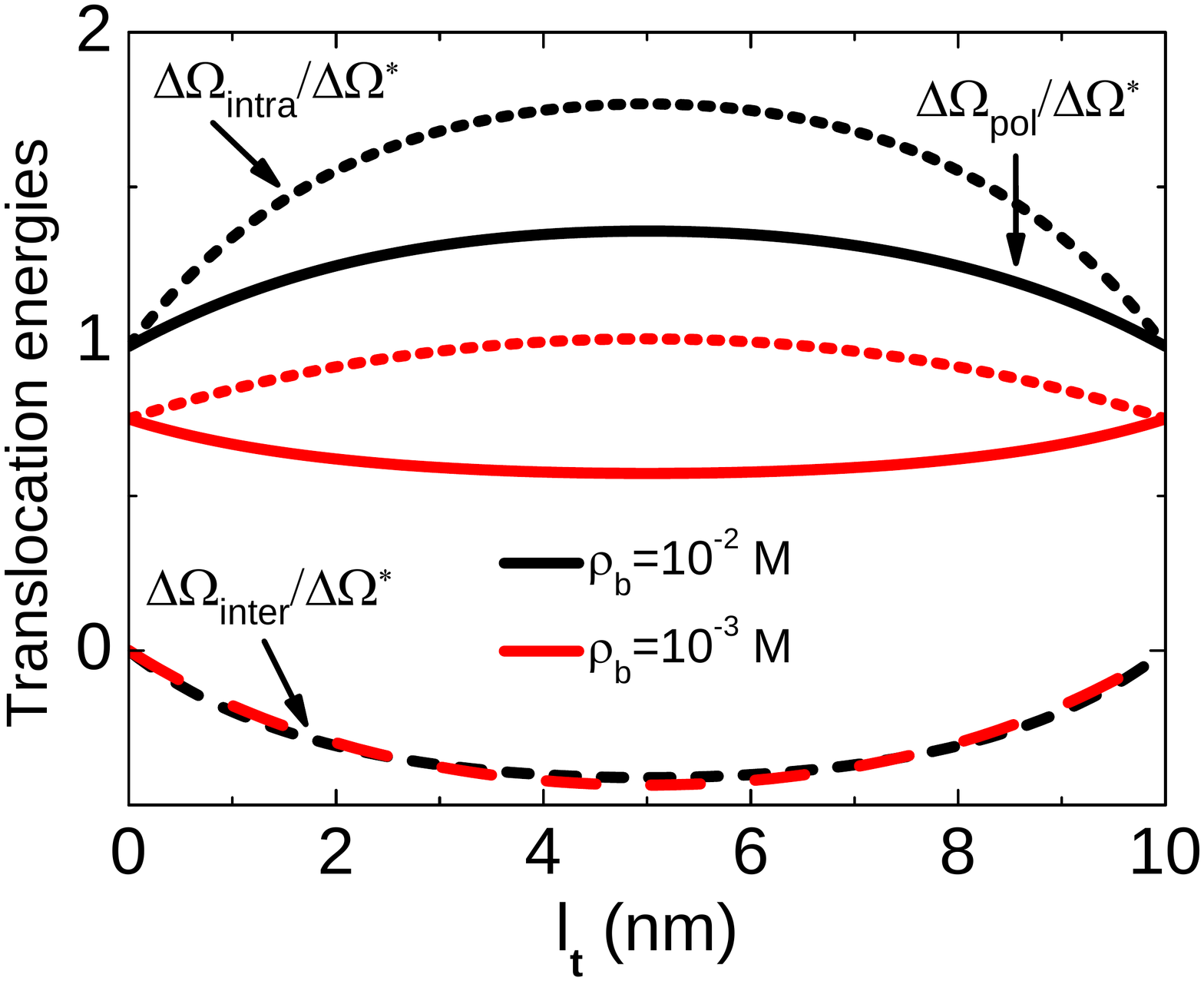}
\caption{(Color online) Electrostatic grand potential of the translocating polymer rescaled by the characteristic energy $\Delta\Omega^*=k_BT\ell_B\tl^2/(2\kappa)$. Dotted curves: Self-energy of Eq.~(\ref{eq15}). Dashed curves: Interaction energy between the {\it cis} and the {\it trans} portions from Eq.~(\ref{eq16}). Solid curves: The total grand potential of Eq.~(\ref{eq9}). Bulk salt density: $\rho_b=10^{-2}$ M (black curves) and $\rho_b=10^{-3}$ M (red curves). The remaining parameters are the same as in Fig.~\ref{fig2}.}
\label{fig4}
\end{figure}

The appearance of an attractive well despite the presence of strongly repulsive image-charge interactions may at first seem counterintuitive. In order to probe the physical mechanism behind this peculiarity, in Fig.~\ref{fig4}, we plot the grand potential components~(\ref{eq15}) and~(\ref{eq16}) rescaled by the characteristic energy $\Delta\Omega^*=k_BT\ell_B\tl^2/(2\kappa)$ at two different salt densities. First, the profile of the DNA self-energy $\Delta\Omega_{\rm intra}(l_t)$ induced by image-charges (dotted curves) is seen to be convex up at all salt densities, thus driving the polymer away from the {\it trans} side. Then, the purely 
negative {\it trans-cis} interaction energy $\Delta\Omega_{\rm inter}(l_t)$ of Eq.~(\ref{eq16}) (dashed curves) exhibits a convex down shape. Hence, this contribution attracts the right half of the polymer towards the {\it trans} side. The negative sign of the interaction term $\Delta\Omega_{\rm inter}(l_t)$ results from the fact that the dielectric mismatch prevents the electric field lines from penetrating into the membrane. This in turn reduces the strength of the electrostatic coupling between the {\it cis} and the {\it trans} portions. At the salt density $\rho_b=10^{-2}$ M (black curves), the repulsive self-energy dominates the membrane-induced attractive 
{\it trans-cis} interaction which results in a total grand potential $\Delta\Omega_{\rm pol}(l_t)$ of convex up shape. Reducing the salt density to $\rho_b=10^{-3}$ M (red curves), the rescaled self-energy $\Delta\Omega_{\rm intra}(l_t)/\Delta\Omega^*$ is significantly lowered while the rescaled {\it trans-cis} interaction energy $\Delta\Omega_{\rm inter}(l_t)/\Delta\Omega^*$ barely changes. As a result, in this dilute salt regime the {\it trans-cis} interaction takes over the DNA self-interaction and the total polymer grand potential $\Delta\Omega_{\rm pol}(l_t)$ acquires a convex down shape, favouring the translation of the DNA towards the {\it trans} side.

We can conclude that the lower the salt concentration, the stronger the contribution of the attractive {\it trans-cis} 
interaction with respect to the repulsive image-charge contribution.  We will now scrutinize the competition between these two effects in terms of the electrostatic force exerted by the dielectric membrane on the translocating polyelectrolyte. 
According to Eq.~(\ref{eq9}), the net electrostatic force on the DNA $F_{\rm pol}(l_t)=-d\Delta\Omega_{\rm pol}(l_t)/d l_t$ 
can be decomposed as $F_{\rm pol}(l_t)=F_{\rm intra}(l_t)+F_{\rm inter}(l_t)$. The force components corresponding to the the grand potentials~(\ref{eq15}) and~(\ref{eq16}) are given by
\bea
\label{eq21}
\frac{F_{\rm intra}(l_t)}{k_BT\ell_B\tl^2}&=&-\mathrm{Ei}(-2\kappa l_t)+\mathrm{Ei}(-\kappa l_t)\\
&&-\mathrm{Ei}\left[-\kappa(L-l_t)\right]+\mathrm{Ei}\left[-2\kappa(L-l_t)\right];\nonumber\\
\label{eq22}
\frac{F_{\rm inter}(l_t)}{k_BT\ell_B\tl^2}&=&\mathrm{Ei}\left[-\kappa(d+L-l_t)\right]-\mathrm{Ei}\left[-\kappa(d+l_t)\right].\nonumber\\
\eea
In Fig.~\ref{fig4}, the characteristic salt density where the total polymer grand potential switches from convex to concave corresponds to the point where the total electrostatic force at $l_t=0$ turns from negative to positive, i.e. $F_{\rm pol}(0)=F_{\rm intra}(0)+F_{\rm inter}(0)>0$. Taking the limit of long polymers $\kappa L\gg1$ for the sake of simplicity, from Eqs.~(\ref{eq21})-(\ref{eq22}), we find that this condition is satisfied if $-\mathrm{Ei}(-\kappa d)<\ln(2)$, or
\be
\label{eq23}
\kappa\lesssim\frac{0.4}{d}.
\ee
Interestingly, the inequality in Eq. (\ref{eq23}) indicates that the thicker the dielectric membrane, the lower 
the critical salt concentration where the half-translocation state switches from unstable to metastable. This stems from the fact that the thickness of the membrane amplifies the repulsive image-charge effect. For the parameters in 
Fig.~\ref{fig4}, this characteristic salt density is $\rho_b\approx0.004$ M. In the pure solvent limit $\kappa\to0$, the image-charge induced force of Eq. (\ref{eq21}) vanishes, $F_{\rm intra}=0$, and the {\it trans-cis} coupling force of Eq. (\ref{eq22}) reads
\be\label{eq24}
F_{\rm inter}(l_t)=\frac{k_BT}{\ell_B}\ln\left(\frac{d+L-l_t}{d+l_t}\right),
\ee
where we took into account the Manning limit~(\ref{man}). We note that as the DNA penetrates the pore,  i.e. for $0\leq l_t\leq L/2$, the force from Eq. (\ref{eq24}) having a positive value is directed to the mid-pore. Hence,  in pure solvents, the electrostatics of the translocation phase is solely governed by the attractive {\it trans-cis} interaction force. 
\begin{figure}
\includegraphics[width=1.1\linewidth]{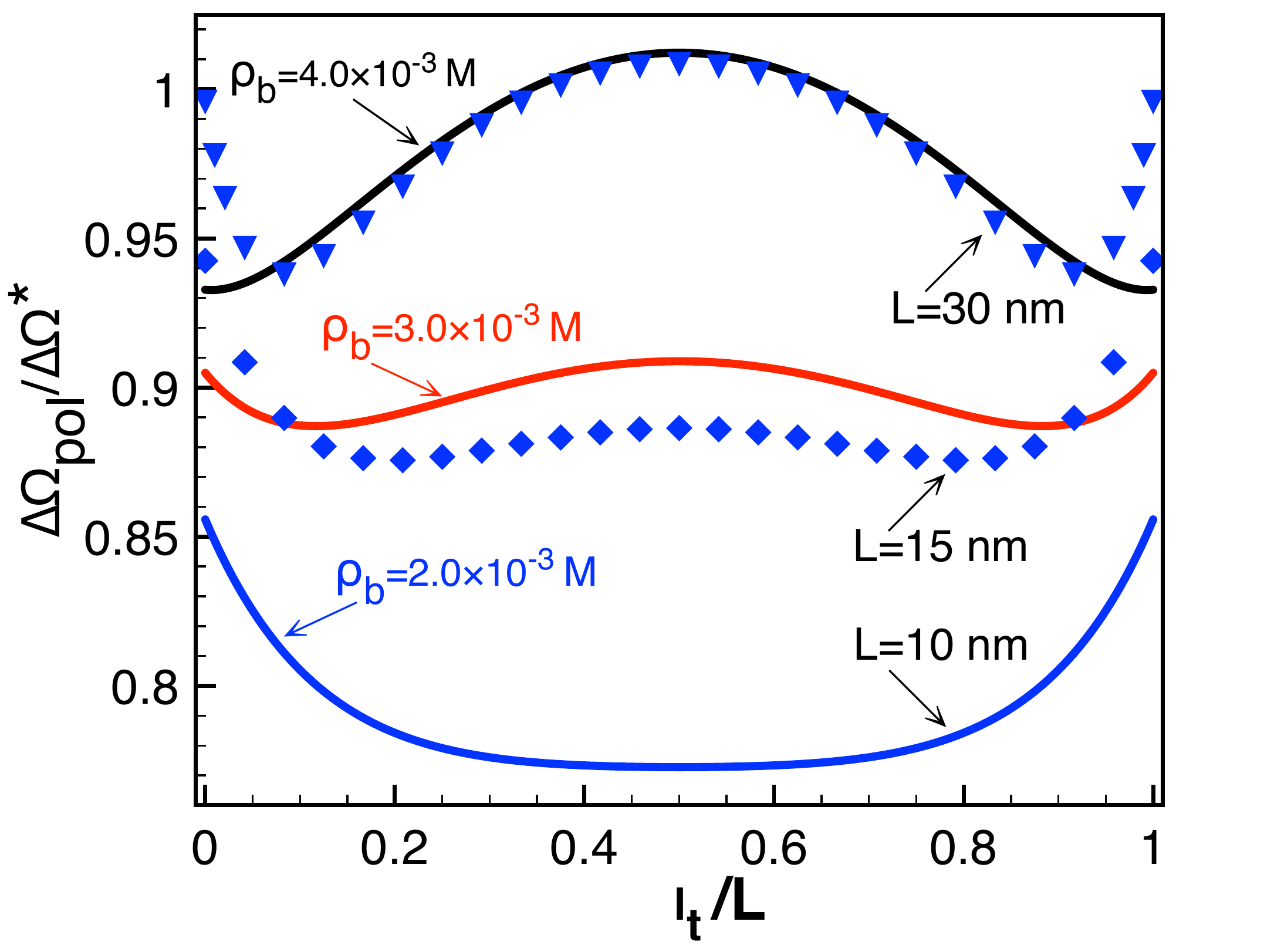}
\caption{(Color online) Electrostatic grand potential of the translocating polymer rescaled by the characteristic energy $\Delta\Omega^*=k_BT\ell_B\tl^2/(2\kappa)$ versus the adimensional translocation coordinate $l_t/L$. Salt densities are $\rho_b=0.002$ M (blue curves and symbols), $\rho_b=0.003$ M (red curve), and $\rho_b=0.004$ M (black curve). Polymer lengths are $L=10$ nm (solid curves), $L=15$ nm (diamonds), and $L=30$ nm (triangles). The remaining parameters are the same as in Fig.~\ref{fig2}.}
\label{fig5}
\end{figure}

\subsection{Polymer length} 
\label{lh}

We now scrutinize the influence of the DNA length on the translocation phase. To this end, we focus on the evolution of the grand potential in the transition regime of Fig.~\ref{fig4} where the grand potential switches from convex down to up. 
Figure \ref{fig5} illustrates the rescaled grand potential profile of the translocating DNA at various salt densities and polymer lengths. We now set the polymer length to $L=10$ nm (solid curves). As we gradually increase the bulk electrolyte concentration from $\rho_b=0.002$ M (blue curve) to $0.004$ M (black curve) with the half-translocated state turning from metastable to unstable, at the ion concentration $\rho_b=0.003$ M (red curve), the system passes through an intermediate state, where the grand potential exhibits two minima at 
finite {\it trans} and {\it cis} lengths. The presence of these two metastable minima may result in oscillations of a translocating polymer between the {\it cis} and the {\it trans} sides of the membrane.

Next we set the salt density to $\rho_b=0.002$ M and change the polymer length $L$ (blue curve and symbols). The comparison of the symbols and curves indicates that the increase of the polymer length is qualitatively equivalent to an increase in
 the ion density. Namely, the metastable half-translocated state at length $L=10$ nm (solid blue curve) becomes unstable for the longer polymer length $L=15$ nm (diamonds), with the appearance of two minima at finite translocation lengths. With a further increase of the polymer length to $L=30$ nm (triangles), the translocation barrier at $l_t=L/2$ increases 
 and the metastable minima split farther. In other words, the membrane is more repulsive to longer polymers. This stems from the fact that an increase in the polymer length amplifies the relative weight of the repulsive self-energy $\Delta\Omega_{\rm intra}(l_t)$ with respect to the attractive {\rm trans-cis} 
 coupling energy $\Delta\Omega_{\rm inter}(l_t)$. In the following section we evaluate the effect of the membrane charge on this competition.

\subsection{Membrane charge}
\label{char}

\begin{figure}
\includegraphics[width=1.1\linewidth]{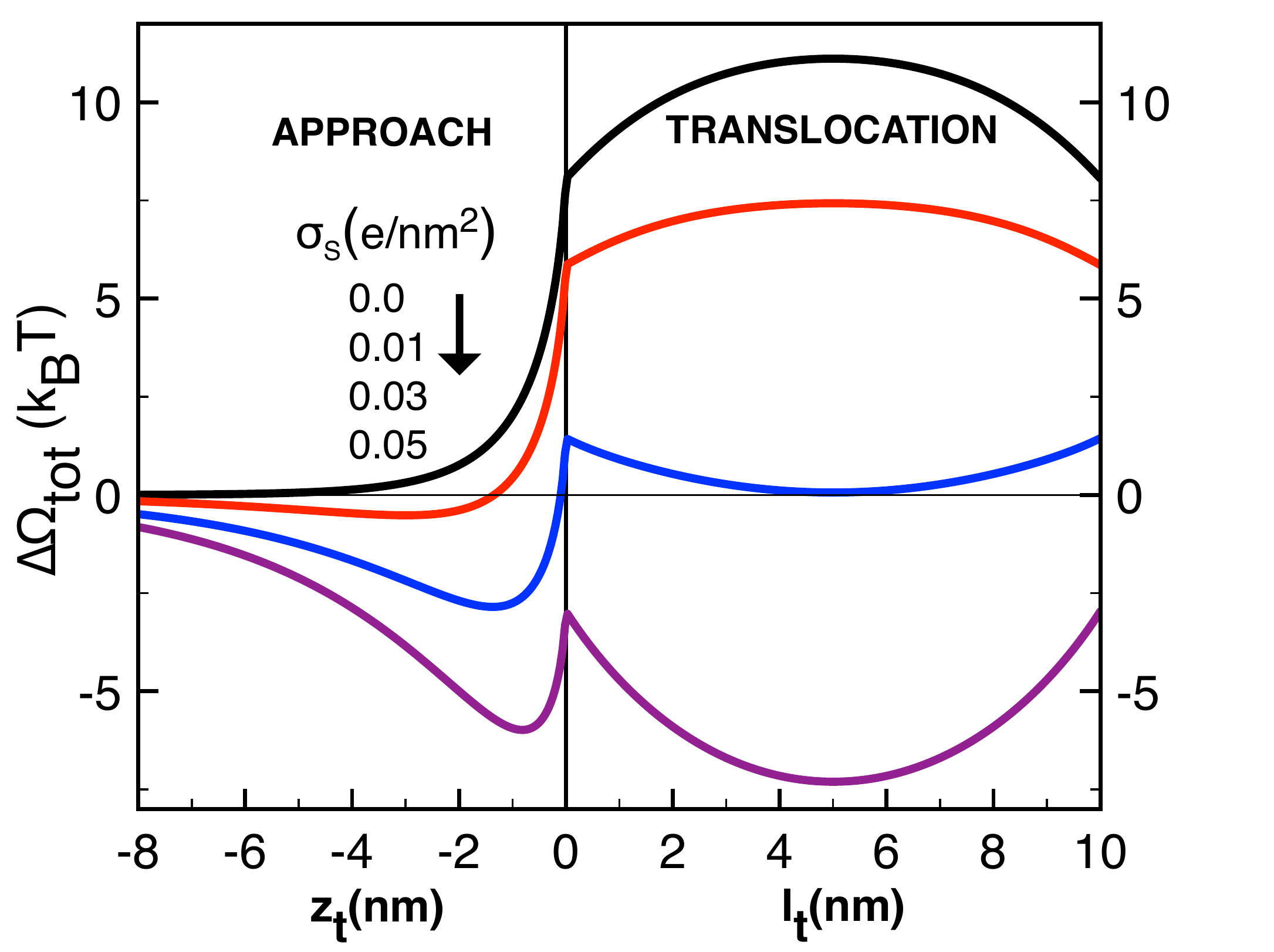}
\caption{(Color online) Polymer grand potential at the permittivity $\e_m=0.0$ for various membrane charges. The salt density is $\rho_b=0.01$ M. The remaining parameters are the same as in Fig.~\ref{fig2}.}
\label{fig6}
\end{figure}

We consider here translocating polymers through charged membranes with permittivity $\e_m=0$. Figure \ref{fig6} illustrates the total grand potential $\Delta\Omega_{\rm tot}$ of Eq.~(\ref{eq0}) at various membrane charges. The grand potential of the approaching polymer is obtained from Eqs.~(\ref{eq3II}) and (\ref{eq4}) in the form
\bea
\label{ap}
\frac{\Delta\Omega_{\rm tot}(z_t)}{k_BT}=\frac{\ell_B\tl^2}{2\kappa}G(z_t)-\frac{2Q_{\rm eff}(L)}{\mu\kappa}e^{\kappa z_t}.
\eea
The grand potential of the translocation phase follows from Eq.~(\ref{eq14II}) and Eqs.~(\ref{eq15})-(\ref{eq16}) as
\bea
\label{tr}
\frac{\Delta\Omega_{\rm tot}(l_t)}{k_BT}&=&\frac{\ell_B\tl^2}{2\kappa}H(l_t)-\frac{\ell_B\tl^2}{\kappa}F(l_t)\\
&&-\frac{2}{\mu\kappa}\left[Q_{\rm eff}(l_t)+Q_{\rm eff}(L-l_t)\right].\nonumber
\eea
We set the salt density to $\rho_b=0.01$ M where the neutral membrane is purely repulsive (black curve). Increasing the membrane charge to $\sigma_m=0.01$ $e/\mbox{nm}^2$ (red curve), the translocation barrier of the neutral membrane survives but an attractive minimum close to the membrane surface takes place at $z_t\approx-2$ nm. Thus, in weakly charged membranes, the DNA should be trapped in the vicinity of the membrane wall. At a stronger membrane charge of $\sigma_m=0.03$ $e/\mbox{nm}^2$ (blue curve), the attractive minimum becomes deeper while the translocation barrier becomes a metastable well. Finally, at the largest charge density $\sigma_m=0.05$ $e/\mbox{nm}^2$ considered in Fig.~\ref{fig6}, this situation is reversed as the minimum outside the membrane turns to metastable and the half-translocated state $\l_t=L/2$ becomes a stable energy minimum. 

\begin{figure}
\includegraphics[width=1.15\linewidth]{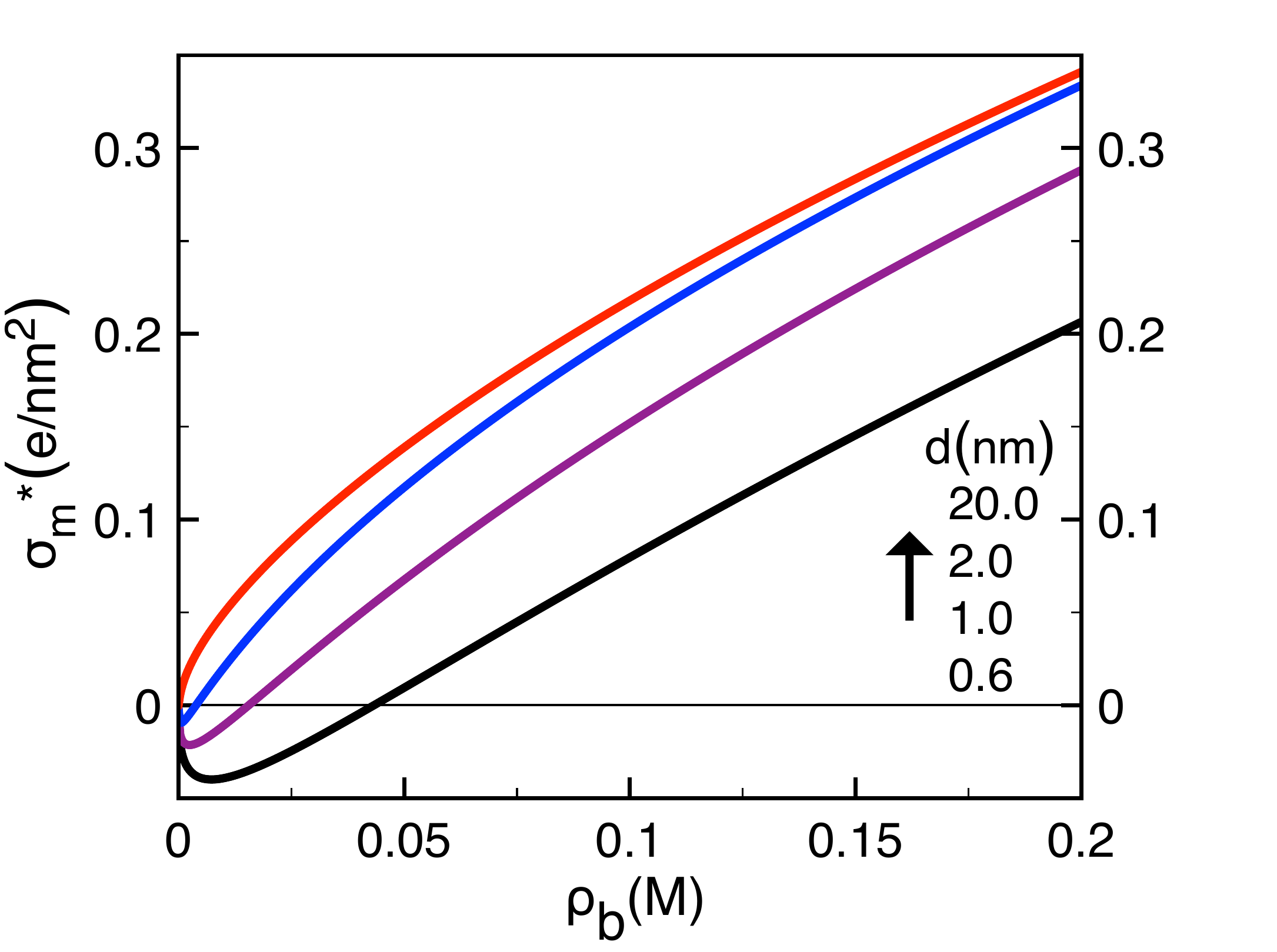}
\caption{(Color online) Critical membrane charge of Eq. (\ref{scr}) versus the bulk salt density for various membrane
thicknesses given in the legend. The areas above and below each curve correspond to the parameter regime where the translocating grand potential is either attractive or repulsive, respectively. The model parameters are the same as in Fig.~\ref{fig2}.}
\label{fig8}
\end{figure}

We have thus found that the variation of the membrane charge distribution over the narrow regime $0\leq\sigma_m\leq0.05$ $e/\mbox{nm}^2$ drastically changes the grand potential landscape and turns the membrane from strongly repulsive to attractive. This suggests that the chemical modification of the membrane surface properties is another factor that allows for a sensitive control on the DNA motion. Motivated by this fact, we calculate next the lowest value of the membrane charge where the translocation barrier at $\ell_t=L/2$ switches to a minimum. We proceed as in Section~\ref{salt} and evaluate the electrostatic force $F_{\rm tot}(l_t)=-d\Delta\Omega_{\rm tot}(l_t)/d l_t$ on DNA from Eq.~(\ref{tr}). Setting the force at the surface to zero, $F_{\rm tot}(l_t=0)=0$, and considering the case of long polymers $\kappa L\gg1$ by taking the limit $L\to\infty$, we get the characteristic charge where the slope of the grand potential curve switches from positive to negative as
\be
\label{scr}
\sigma_m^*=\frac{\kappa\tl}{4\pi}\left[\ln(2)+\mathrm{Ei}(-\kappa d)\right].
\ee
In Eq.~(\ref{scr}), the first positive term in the bracket corresponds to repulsive image charge interactions and increases the critical charge. The latter is in turn attenuated by the negative second term of Eq.~(\ref{scr}) associated with the attractive {\it trans-cis} coupling energy. In Fig.~\ref{fig8}, we plot the characteristic charge of Eq. (\ref{scr}) against the bulk ion density at various membrane thicknesses. The surface above and below each curve corresponds to the parameter range where the membrane is a metastable and an unstable point, respectively. We see that at low salt concentrations, the characteristic membrane charge is negative and drops with increasing salt density until it reaches a minimum. Beyond this turning point, the charge rises monotonically with the salt density and becomes positive.

The non-monotonic behaviour of the characteristic charge curves is due to the competition between image-charge and {\it trans-cis} interaction terms in Eq.~(\ref{scr}). We focus first on the large ion density regime $\kappa d\gtrsim1$. Equation 
(\ref{scr}) shows that at large ion concentrations, the contribution from the attractive {\it trans-cis} coupling is exponentially screened. Consequently, only the repulsive image charge contribution survives in this regime. In addition, the salt screening attenuates the field induced by the membrane charge. Thus, the larger the salt density, the larger the positive membrane charge should be ($\rho_b\uparrow\;\sigma_m^*\uparrow$) in order for the membrane-DNA attraction to dominate the DNA-image charge repulsion. This explains the positive slope of the critical charge curves at large ion concentrations. As the membrane thickness amplifies image-charge effects we also note that at fixed ion density, the larger the membrane thickness $d$, the larger the critical membrane charge ($d\uparrow\;\sigma_m^*\uparrow$). 

We consider now the dilute salt regime $\kappa d\lesssim1$ of Fig.~\ref{fig8} where the critical membrane charge exhibits non-monotonic behaviour. In Section~\ref{salt} we found that in weak electrolytes,  the attractive {\it trans-cis} interaction takes over the repulsive image-charge effect. Thus, as the neutral membrane is already attractive to the DNA, one needs a negative membrane charge for the polymer-membrane charge repulsion to compensate the attractive {\it trans-cis} coupling energy, explaining the negative sign of the characteristic charge in Fig.~\ref{fig8}. Indeed, expanding Eq.~(\ref{scr}) for $\kappa d\ll1$ we get
\be
\label{scr2}
\sigma_m^*\approx\frac{\kappa}{4\pi\ell_B}\left[\gamma+\ln(2\kappa d)\right],
\ee
which is negative since the logarithmic term is strongly negative for $\kappa d\ll1$. In Eq.~(\ref{scr2}), $\gamma\approx0.577(2)$ stands for the Euler gamma function~\cite{math} and we also took into account  the Manning limit of the polymer charge~(\ref{man}). By differentiating Eq.~(\ref{scr2}) with respect to $\kappa$ and setting the result to zero, the position of the minimum of $\sigma_m^*$ follows as
\be
\label{kcr}
\kappa_c\approx\frac{e^{-\gamma-1}}{2d}.
\ee 
Substituting Eq.~(\ref{kcr}) into Eq.~(\ref{scr2}), the minimum of the critical surface charge reads
\be
\label{scr3}
\sigma_m^*(\kappa_c)\approx-\frac{e^{-\gamma-1}}{8\pi\ell_Bd}.
\ee
In agreement with Fig.~\ref{fig8}, Eqs.~(\ref{kcr}) and~(\ref{scr3}) indicate that the larger the membrane thickness, the weaker the minimum membrane charge and the corresponding salt density where the turning point takes place. The complex behaviour of the phase diagram in Fig.~\ref{fig8} embodied by the simple relation~(\ref{scr}) calls for an experimental verification.

\section{Conclusions and Summary}

In this work, we have developed the first electrostatic model of stiff polyelectrolyte translocation through dielectric membranes in electrolyte solutions. The theory can fully account for the DNA-image-charge and DNA-membrane charge interactions beyond MF level. Unlike previous electrostatic formalisms that considered exclusively the portion of the DNA located inside the pore~\cite{e3,Buyuk2014},  the model can take into account electrostatic interactions associated with the DNA segments 
in the {\it trans} and the {\it cis} sides of the membrane. This becomes crucial for translocation experiments with graphene-based membranes whose thickness can be lowered up to $d\approx6$ {\AA}~\cite{Garaj}. By introducing a charge renormalisation procedure applied to the polyelectrolyte, we have also 
been able to overcome the DH approximation of Ref.~\cite{Buyuk2016}. In translocation events through neutral membranes, we have shown that the dielectric mismatch between the membrane and the solvent plays a leading role. Due to the resulting image-charge effects, at large ion densities $\rho_b\gtrsim0.01$ M, polymers translocating C-based membranes with low permittivity $\e_m\approx2$ experience a large repulsive barrier of $\approx10$ $k_BT$. In engineered membranes with large permittivity $\e_m>\e_w$, the membrane becomes in turn strongly attractive as the translocation grand potential exhibits a minimum of the order of about $10$ $k_BT$.  

In the most relevant case of low-permittivity neutral membranes, translocation is driven by competition between repulsive DNA-image charge interactions excluding the polymer from the membrane, and the coupling between the {\it trans-cis} portions of the DNA molecule attracting the latter towards the {\it trans} side. The attractive force is due to the dielectric membrane that prevents the electric field lines originating from the {\it trans} and the {\it cis} portions to pass to the other side of the membrane volume. This mechanism weakens the electrostatic coupling between these portions, reducing the DNA grand potential with respect to the bulk liquid. In dilute salt solutions with density $\rho_b\lesssim0.01$ M or for short polymer sequences, the attractive {\it trans-cis} coupling dominates the repulsive image-charge-induced barrier. As a result, the membrane becomes a metastable attraction point that is expected to trap the translocating DNA over considerable time intervals. This peculiarity is the key result of our work. Since an accurate sequencing of DNA requires control and reduction of the DNA translocation velocity~\cite{e5}, our result suggests that this can be achieved most simply by tuning the salt concentration of the solution.

In weakly charged membranes, the competition between the image-charge repulsion and the membrane-charge attraction results in an attractive well close to the membrane surface at $z_t\approx-1$ nm. At the surface charge $\sigma_s\approx0.01$ $e/\mbox{nm}^2$, this attractive minimum is followed by a repulsive translocation barrier at $l_t=L/2$. Thus, polymers approaching weakly charged membrane interfaces should be trapped outside the membrane. At stronger membrane charges, the attractive well becomes metastable while the translocation barrier switches to a stable minimum of the potential landscape, driving the polymer to the {\it trans} side. This mechanism presents itself as an alternative way to control DNA-membrane interactions via the chemical modification of the membrane surface properties or by tuning the pH of the solution.

For the sake of analytical simplicity and physical transparency, there are several approximations in the polyelectrolyte-membrane system considered here. First, our rigid polyelectrolyte model does not account for configurational fluctuations of 
the DNA molecules. This limitation can in principle be overcome by coupling the Coulomb liquid model with Edward's path integral formulation of fluctuating polymers~\cite{dun}. However, in the most relevant case of C-based membrane, the large persistence length of ds-DNA molecules 
$l_p \approx 50$ nm is expected to be enhanced by image-charge forces. Thus, the inclusion of polymer fluctuations is not expected to qualitatively change the conclusions of the present work. Furthermore, as the theory includes the membrane charge at the DH level, we have restricted ourselves to weakly charged membranes in contact with monovalent electrolytes. Future work including charge correlations at the full one-loop level will allow us to consider the case of strongly charged membranes and multivalent ions. Finally, the present theory is based on the assumption of local equilibrium for the calculation of the grand potential. In the future we plan to include an explicit
description of dynamics in polymer translocation.

\acknowledgements{This work has been supported indirectly by the Academy of Finland through its COMP Center of Excellence
Program under project numbers 284621 and 287750. It should have been supported directly, too.}

\smallskip
\appendix
\section{Polymer self-energy}
\label{a1}

In this Appendix, we review the general lines of the derivation of the polymer self-energy. In Ref.~\cite{Buyuk2016}, we showed that at the DH level, electrostatic self-energy of the polymer reads
\be\label{eqA1}
\Omega_{\rm pol}=k_BT\int\frac{\mathrm{d}\br\mathrm{d}\br'}{2}\sigma_p(\br)v_{\rm DH}(\br,\br')\sigma_p(\br'),
\ee
where $\sigma_p(\br)$ is the polymer charge density and the Green's function $v_{\rm DH}(\br,\br')$ is the solution of the DH equation
\be\label{eqA2}
\left[-\frac{1}{\beta e^2}\nabla\cdot\e(\br)\nabla+2\rho_b q^2\right]v_{\rm DH}(\br,\br')=-\delta(\br-\br').
\ee
In Eq.~(\ref{eqA2}), $\beta=1/(k_BT)$ is the inverse thermal energy, $e$ the electron charge, $\e(\br)$ the dielectric permittivity function, $\rho_b$ the bulk ion density, and $q$ stands for the ionic valency. Moreover, for the same membrane-electrolyte geometry, the dielectric permittivity profile is
\be\label{eqA4}
\e(z)=\e_w\theta(-z)+\e_m\theta(z)\theta(d-z)+\e_w\theta(z-d),
\ee
with $\e_m$ being the membrane permittivity and $\e_w=80$ the solvent permittivity. Due to the translational symmetry in the membrane plane, one can Fourier-expand the Green's function as
\be\label{eqA5}
v_{\rm DH}(\br,\br')=\int\frac{\mathrm{d}^2\bk}{4\pi^2}\;e^{i\bk\cdot\left(\br_\pa-\br_\pa'\right)}\tv_{\rm DH}(z,z').
\ee
In Ref.~\cite{Buyuk2016}, the Fourier-transformed solution of Eq.~(\ref{eqA2}) was calculated in the form
\bea
\label{eqA6}
\tv_{\rm DH}(z\leq0,z'\leq0)&=&\tv_b(z-z')\\
&&+\frac{2\pi\ell_B}{p}\frac{\Delta\left(1-e^{-2kd}\right)}{1-\Delta^2e^{-2kd}}e^{p(z+z')}\nonumber\\
\label{eqA7}
\tv_{\rm DH}(z\geq d,z'\geq d)&=&\tv_b(z-z')\\
&&+\frac{2\pi\ell_B}{p}\frac{\Delta\left(1-e^{-2kd}\right)}{1-\Delta^2e^{-2kd}}e^{p(2d-z-z')}\nonumber,
\eea
and
\bea
\label{eqA8}
\tv_{\rm DH}(z,z')&=&\tv_b(z-z')\\
&&+\frac{2\pi\ell_B}{p}\frac{(1-\Delta^2)e^{(p-k)d}+\Delta^2e^{-2kd}-1}{1-\Delta^2e^{-2kd}}e^{-p|z-z'|}\nonumber
\eea
for $z'\leq0$ and $z\geq d$, or $z'\geq d$ and $z\leq0$. In Eqs.~(\ref{eqA6})-(\ref{eqA8}), the dielectric discontinuity function is defined as
\be
\Delta=\frac{\e_wp-\e_mk}{\e_wp+\e_mk}.
\ee
We finally note that in Eqs.~(\ref{eqA6})-(\ref{eqA8}),  we introduced the bulk part of the Fourier-transformed DH potential
\be\label{eqA9}
\tv_b(z-z')=\frac{2\pi\ell_B}{p}e^{-p|z-z'|}.
\ee
By inserting the charge density function of Eq. (\ref{pc}) together with the Fourier expansion~(\ref{eqA5}) into the right-hand-side of Eq.~(\ref{eqA1}), evaluating the integrals over the membrane plane, and subtracting the bulk part due to Eq.~(\ref{eqA9}), the grand potential finally takes the form
\be
\label{eqA10}
\frac{\Delta\Omega_{\rm pol}}{k_BT}=\lambda^2\int_0^\infty\frac{\mathrm{d}kk}{4\pi}\iint_{-\infty}^{+\infty}\mathrm{dz}\mathrm{dz'}g(z)\delta\tv_{\rm DH}(z,z')g(z'),
\ee
where we defined the part of the Fourier transformed Green's function associated with the presence of the membrane as
\be
\label{eqA11}
\delta\tv_{\rm DH}(z,z')=\tv_{\rm DH}(z,z')-\tv_b(z-z').
\ee

\end{document}